\documentclass[usenatbib]{aa}

\usepackage{graphicx}
\usepackage{amssymb}
\usepackage{epsfig}
\usepackage{psfig}
\usepackage{natbib}

\def\chiq{$\chi^2$}

\def\thesource{2S~ 0114+65}
\def\be{\begin{equation}}
\def\ee{\end{equation}}
\def\lsim{\lower0.5ex\hbox{$\; \buildrel < \over \sim \;$}}
\def\gsim{\lower0.5ex\hbox{$\; \buildrel > \over \sim \;$}}
\def\sref#1{\S~\ref{#1}}

\begin{document}

\title{INTEGRAL High Energy  Observations  of  2S 0114+65}
\author{E. W. Bonning\inst{1}\fnmsep\thanks{\email{erin.bonning@obspm.fr}},
M. Falanga\inst{2}}

\offprints{E. W. Bonning}
\titlerunning{{\em INTEGRAL} Observations of 2S 0114+65}
\authorrunning{E. W. Bonning \& M. Falanga }

\institute{ LUTh,  Observatoire de Paris, F-92195 Meudon Cedex, France
\and CEA Saclay, DSM/DAPNIA/Service d'Astrophysique (CNRS FRE
  2591), F-91191, Gif sur Yvette, France
}

\abstract{
We report the first {\em INTEGRAL} timing and spectral analysis of the high
mass X-ray binary source 2S 0114+65 at high energies (5--100 keV). The pulse
period was found at 2.668 hr with a high pulsed fraction, $\sim$80\%
in both the 20--40 keV and 40--80 keV energy bands.  The spin-up trend
over $\sim$8 years was measured to be $\dot P = -8.9 \times 10^{-7}$ s
s$^{-1}$. The hard X-ray  spectrum  
obtained with JEM-X/ISGRI is well described by a high energy exponential
cut-off power law model where the estimated luminosity is 1.8 $\times
10^{36}$ erg s$^{-1}$ in the 5--100 keV energy band, for a source
distance of 7.2 kpc. We tentatively identify a cyclotron
resonance scattering feature at $\sim$22 keV with one harmonic, implying a 
magnetic field of $\sim2.5 \times 10^{12}$ G. 
}
\maketitle

\section{Introduction}
\label{sec:intro}

The X-ray source \thesource\  is a high mass X-ray
binary (HMXB) consisting of an accreting neutron star (NS) and a type
B1a supergiant optical counterpart (LS I+65010) at a
distance of $\sim$7.2 kpc \citep{reig96}. The pulse period was
first measured at 2.78 hr \citep{finley92} with a 11.59 day orbital
period \citep{crampton85}. A spin-up rate over $\sim$11 yr was
observed to be $\sim6.2\times10^{-7}$s s$^{-1}$  (Hall et al. 2000).  
 The long spin period makes this the slowest
known X-ray pulsar. Different models have been suggested 
in the past to explain the peculiar long period pulsation seen in
2S 0114+65 \citep[c.f.][~and references therein]{hall00}. Most recently,
\citet{li99} suggest that such a slow period is
possible if 2S 0114+65 was born as a 
magnetar with an initial magnetic field strength of $\ga$10$^{14}$ G,
decaying to a current value of the order  10$^{12}$ G,
allowing the NS to spin down to the measured spin period within the
lifetime of its companion. 
The magnetic field has not been  measured up to now, as no cyclotron
lines have been detected in the X-ray spectrum up to 20 keV.
Recently, \citet{farrell05} found evidence for a super-orbital 
period of 30.7 days, suggesting the presence of warped, precessing
accretion disk. 

Observations by \citet{hall00} with  {\em RXTE} show that the 3--20
keV spectrum is best fit by an absorbed power law with exponential high energy
cut-off, with photon index $\sim$1.4, cut-off energy $\sim$8 keV, and
folding energy $\sim$20 keV, being typical parameters for a HMXB. They
find evidence for a fluorescent iron line when 
the source is in a low emission state, consistent with measurements of
\citet{yamauchi90}.

In this paper, we report the results of high energy {\em INTEGRAL}
observations of \thesource. The observations are described in
\sref{sec:integral}. Imaging, timing, and spectral analyses are  presented in 
\sref{sec:res}. A discussion of the results is given in
\sref{sec:conclusion}.   

\section{Observations and Data Analysis}
\label{sec:integral}

The present dataset was obtained during the Target of Opportunity
 (ToO) A02  {\em INTEGRAL}  \citep{w03}  observation of the Cas A
 region performed  from 5--6 and 7--9 December 2004 ( 53345.6
 -- 53346.8 and  53347.8 -- 53349.8 MJD), i.e. from part of
{\em INTEGRAL} satellite revolutions  262 and 263. 
We use  data from the coded mask imager IBIS/ISGRI \citep{u03,lebr03}
for a total exposure of 181.9 ks, and from the JEM-X monitor
 \citep{lund03} for a total exposure time of 11 ks.     
For ISGRI, the data were extracted for all pointings with a
source position offset $\leq$ $12^{\circ}$, and for JEM-X with an
offset $\leq$ $3.5^{\circ}$.
The spectrometer (SPI) was not used to extract the hard X-ray spectrum
due to the lower sensitivity of this instrument with respect to
IBIS/ISGRI for a weak source below 100 keV. Above 100 keV, 2S 0114+65
was neither consistently detected in a single exposure nor in the total
exposure time. 
Data reduction was performed using the standard Offline Science
Analysis (OSA) version 4.2 \citep{c03}.
The background subtracted ISGRI light curve for the timing analysis  
was reduced using  dedicated software $ii\_light\_extract$ (version to
 be included in OSA 5.0).   
The ISGRI light curve per energy band was obtained for each 300 s
long time bin.

\section{Results}
\label{sec:res}

\subsection{ISGRI Imaging}

\begin{figure}
\centerline{\epsfig{file=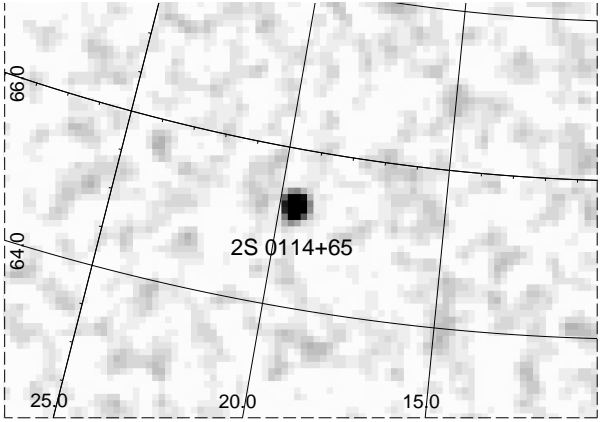,width=8cm}}
\caption
{
The 20--40 keV IBIS/ISGRI mosaicked and deconvolved sky image of the
$\sim182$ ks observation.
Image size is $\sim15^{\circ} \times 5^{\circ}$, centered at 2S 0114+65
position. The pixel size is 5$'$.   2S 0114+65 was
detected at a significance of  $\sim37~\sigma$.
}
\label{fig:ibis_img}
\end{figure}

Fig. \ref{fig:ibis_img} shows a significance map around the source 
 2S 0114+65 in the 20--40 keV energy range. Single pointings were 
deconvolved and analyzed separately, and then combined in mosaic images.
The source is clearly detected at a significance level of
$36.6~\sigma$. In the energy range 40--80 keV, the significance level was
$10.8~\sigma$. 
From the ISGRI data,   2S 0114+65 was
observed with the imaging procedure at the  position 
$\alpha_{\rm J2000} = 01^{\rm h}18^{\rm m}02\fs55$ and $\delta_{\rm J2000} =
65{\degr}17\arcmin02\farcs3$. The source position offset
with respect to the catalog position \citep{bradt83} is 0.5$'$. This is
 within the 90\%   confidence level assuming the imaging source
 location error  given by \citet{gros03}.  

\begin{figure}
\centerline{\epsfig{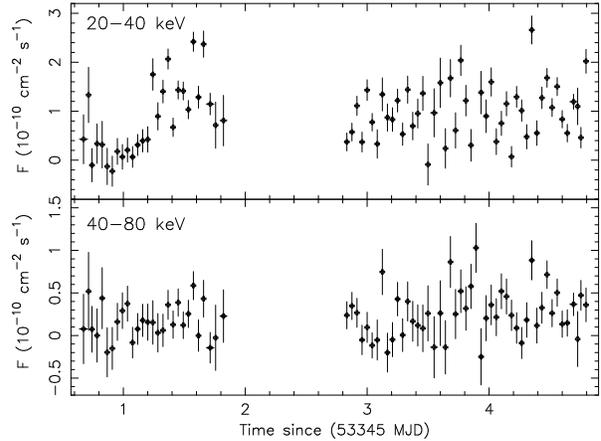}}
\caption
{ {\em INTEGRAL}/ISGRI light curve of \thesource\ in the 20--40 and
40--80 keV energy bands. 
The ISGRI  data are taken from the whole observation
and have been converted to flux assuming a power law  with
an exponential high-energy cut-off model (see Table \ref{table:fit}).
Each point corresponds to a $\sim$2.2 ks interval.
}
\label{fig:lc}
\end{figure}

\subsection{Timing Analysis}
\label{sec:timing}

The ISGRI 20--40 keV and 40--80   keV high energy light curves
were extracted from the images using all available pointings
and are shown in Fig. \ref{fig:lc}. During the first $\sim$30 ks of
the observation, the flux rises to a mean count rate in the 
20--40 keV energy band of $\sim$1.6 cts s$^{-1}$
($\sim$9.5$\times$10$^{-11}$erg s$^{-1}$ cm$^{-2}$). Similarly, but less 
significantly, the flux in the  40--80   keV energy bands attains a
count rate of $\sim$0.5 cts s$^{-1}$ ($\sim$2.6$\times$10$^{-11}$erg
s$^{-1}$ cm$^{-2}$). The count rate is converted to flux  assuming a
power law  with an exponential high-energy cut-off power-law model (see Table
\ref{table:fit}, Section \ref{sec:spectrum}).
The hardness ratio of these energy bands as a function of time
indicates that no significant spectral variability was detected.
  
We searched for coherent pulsations of the source in the 20--40 keV energy   
band where we have the best statistics.    
Using the 300 s binned light curve of the source in the 20--40 keV energy   
band, we computed a Power Density Spectrum (PDS) in the frequency 
range between   
6.5$\times$10$^{-6}$ and 1.6$\times$10$^{-3}$ Hz from Fast Fourier
Transforms. In the   
resulting PDS, an evident signal is present at  $\nu = 1.04\times10^{-4}$
Hz.  The highest peak in the power spectrum
 corresponds to a nominal period of 2.67 hr.
The accurate period was found using the phase-delay  
fitting method. We divided the data over 8 intervals of $\sim23$ ks
each and folded the light curve corresponding to each of the intervals 
using the nominal pulse period. The pulse period was not significantly
detected in the first interval.  \citet{hall00} suggest that
\thesource\ is an eclipsing system, which may explain 
the very low source flux in this first interval. However, using their
ephemeris, the orbital period 11.63 $\pm$ 0.007 days given by
\citet{corbet99}, and  no orbital period evolution, we find
the region of low flux in our observation to 
lie directly within the error of the expected location of the eclipse.
 For intervals 2--8,
we found the pulse phase delays of 
each of the folded light curves with respect to the second interval.
We fitted these phase delays with the first two terms of the equation
$\delta \phi = \delta \phi_{\circ} + \delta\omega(t-t_{\circ})
+\frac{1}{2}\dot\omega(t-t_{\circ})^{2}$, where $\delta
\phi$ is the measured pulse phase delay, $t_{\circ}$  is the mid time of the
observation, $\delta \phi_{\circ}$ is the phase delay at $t_{\circ}$,
$\delta \omega$ is the deviation from the nominal pulse frequency, and
$\dot\omega$ is the pulse frequency derivative. 
The best determination of the mean pulse period was found to be P = 2.668
$\pm$ 0.004 hr.  Errors are at 1$~\sigma$ confidence level. The short
observation time and low statistics did not allow us to determine or
give an upper limit on  the pulse derivative. 

\begin{figure}[t]
   \centering  
   \includegraphics[width=6.0 cm, angle= -90]{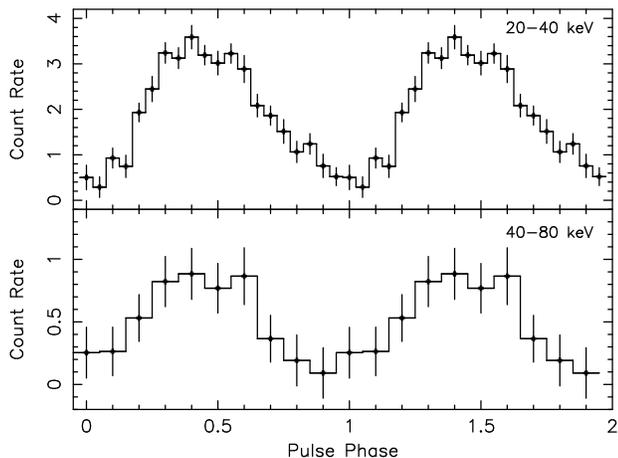}  
 \caption{IBIS/ISGRI background subtracted light curve of \thesource\
  folded  at the best found  NS spin period (2.668 hr). The epoch
  $T_0$ is  arbitrary. From top to bottom the energy ranges are 
 20--40 keV  and 40--80 keV, respectively. The pulse profile is
  repeated once for clarity.} \label{fig:bestperiod}  
 \end{figure}

The folded light curve (excluding the
first interval where the pulse period was not detected above
3$~\sigma$) is shown in Fig. \ref{fig:bestperiod}. The pulse profiles show
a single peak up to 80 keV similar to that observed
previously at lower energies, c.f. \citet{hall00} and
\citet{corbet99}. The pulsed fraction (PF),     
 defined here as PF = $(I_{max}-I_{min})/(I_{max}+I_{min})$, is
 measured at 76 $\pm$5\% in the 20--40 keV energy range and 84
 $\pm$20\% in the 40--80 keV band. 
  
\begin{figure}[t]
   \centering  
  \includegraphics[width=7.3 cm, angle=0]{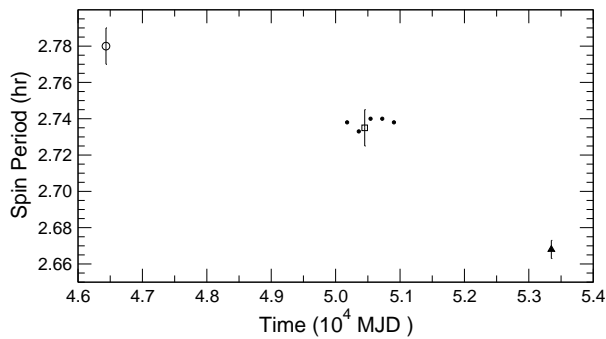}  
 \caption{Spin history of 2S 0114+65. The filled triangle is from this
   work.  The open circle is from  \citet{finley92}, filled circles
   from \citet{corbet99}, and open square from \citet{hall00}.
}
\label{fig:spin_evolution}  
 \end{figure}   

\subsection{Spectral Analysis}
\label{sec:spectrum}

The spectral analysis was done using XSPEC version 11.3 \citep{arnaud96},
combining the 20--100 keV ISGRI data with the
simultaneous 5--20 keV JEM-X data. Due to the short exposure time of
JEM-X, and therefore lower statistics, we
rebinned this data in a 5 channel energy response matrix in order to
constrain the ISGRI spectrum at higher energy.  For ISGRI we used a 15 channel
energy response matrix.  
A constant factor was included in the fit to take into account the
uncertainty in the cross-calibration of the instruments. The factor
was fixed at 1 for the ISGRI data. 
A systematic error of 2\% was applied to the JEM-X/ISGRI spectra which
corresponds to the current uncertainty in the response matrix.
All spectral uncertainties in the results are given at a 90\%
confidence level for single parameters.

The joint JEM-X/ISGRI (5--100 keV)
spectrum  was first fitted with a simple model consisting
of a power law, which was found inadequate with {\chiq}/dof=49.5/17.
In order to compare with previously reported measurements
\citep{hall00}, we fit a common model for HMXBs, namely a power law with
exponential high energy  cut-off. The photon flux density adopts the form
$f(E) = A E^{-\Gamma}e^{(E_{\rm cut}-E)/E_{\rm fold}}$, where $A$ is a
normalization constant. This model gives a better fit with 
{\chiq}/dof=14.5/15, and the resulting  best fit parameters
$\Gamma = 1.6$, $E_{\rm cut} = 9.0$ keV and $E_{\rm fold} = 22.1$ keV,
are consistent within their errors with the  previously published
values of \citet{hall00}.

\begin{figure}
\centerline{\epsfig{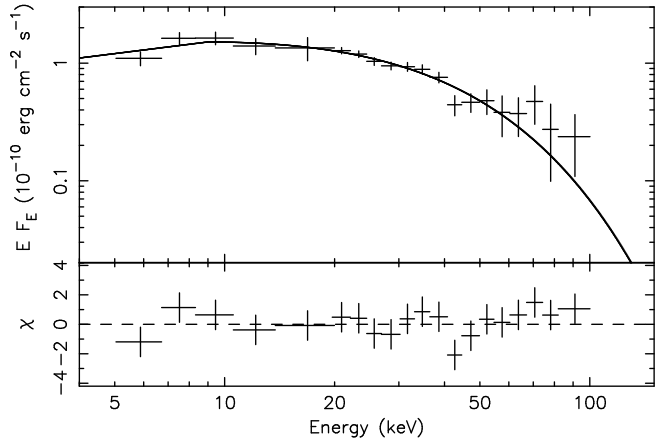}}
\caption
{{\em INTEGRAL}/JEM-X (5--20 keV) and ISGRI (20--100 keV) unfolded spectra of
      \thesource\ with the best-fit power law and exponential 
      high-energy cut-off model. Residuals between the data and model
      are shown in the bottom panel in units of sigma.}
\label{fig:powlaw}
\end{figure}

We note that there is a feature in the residuals to this fit (see
Fig. \ref{fig:powlaw}) around 40--50 keV.  This is suggestive of a
cyclotron resonance scattering feature (CRSF), which is observed in many
accreting X-ray pulsars \citep{coburn02} at high energies. We
attempted to test the hypothesis of a cyclotron feature in the 
spectrum by modifying the
power-law  model with a two-harmonic cyclotron absorption line
\citep{mihara90}. This fit shows a CRSF at
22 keV with a first harmonic at 44 keV and line widths 9.8 and
1.7 keV respectively. Although this model removes the dip feature in
the residuals, the {\chiq}/dof of 5.4/10 does not improve the fit.   

Absorption by neutral hydrogen was not included in these models since previous
reported values of the hydrogen column density, $\sim3 
\times 10^{22}$ cm$^{-2}$ \citep{hall00} did not have effects above 5
keV. Similarly to the timing analysis, we excluded the first 14
pointings of the total ISGRI spectrum.
The best fit parameters of the model together with their errors are
reported in Table \ref{table:fit}. The flux was converted to
luminosity assuming a distance of 7.2 kpc \citep{reig96}.
The fairly large errors reported for the best fit parameters $E_{\rm
  cut}$ and the related $E_{\rm fold}$ are
substantially due to the low statistics of the JEM-X data. 
The unfolded spectrum and the residuals of the data  are
shown in Fig. \ref{fig:powlaw}.

\begin{table}
\caption{Spectral best fit parameters.}
\begin{center}
\begin{tabular}{ll}
\hline 
Dataset  &  \multicolumn{1}{c} {JEM-X/ISGRI}   \\
Model           & {\sc PL} $\times$ {\sc Highecut} \\
\hline 
$\Gamma$               & 1.6$^{+0.5}_{-0.5}$   \\
$E_{\rm cut}$ (keV)    & 9.0$^{+14.1}_{-8.8}$  \\
$E_{\rm fold}$ (keV)   & 22.1$^{+12.1}_{-6.0}$ \\
$\chi^{2}/{\rm dof}$   & 14.5/15            \\
$L_{\rm 5-100}$ keV (erg s$^{-1}$)  & 1.8 $\times 10^{36}$\\
\hline
\end{tabular}
\end{center}

\label{table:fit}
\end{table}

\section{Conclusions} 
 \label{sec:conclusion}
 
We have carried out a temporal and spectral analysis for the first time with
{\em INTEGRAL}  data of the HMXB  \thesource. We determined the spin period
to be 2.668 hr, which indicates that the long-term spin-up of the
NS is still continuing. In Fig. \ref{fig:spin_evolution}, we
show the detailed pulse period history of the source from the first
determination in 1986 with {\em EXOSAT}. 
Since 1996, the spin
period has decreased from $\sim$2.73 hr to $\sim$2.67 hr.  The measured 
spin period derivative from MJD 50451 to MJD 53348 is $\dot P =
-8.9 \times 10^{-7}$ s s$^{-1}$ which is $\sim$1.4 times as large as
the value reported by \citet{hall00}. 
The increase in $\dot P$  likely occurs from an increase of accreting
matter, and shows that the spin-up is not a linear trend.  
The pulsed fraction is observed to be significant up to 80 keV,
76\% in the range 20--40 keV and 84\% in the range 40--80
keV. The detection of pulsed emission up to 80 keV and the measured spin-up
trend strongly support the conclusion that \thesource\ is a rotating neutron
star where the pulsed photons arise from the polar cap on the NS surface.

The energy spectrum of  \thesource\ from 5--100 keV is well
fit by the standard model used for X-ray binary pulsars, i.e. a
power law with  exponential high energy cut-off. Best fit
parameters give a photon
index $1.6$, cut-off energy $9.0$ keV, and folding energy $22.1$
keV. The X-ray luminosity calculated from the model flux is $1.8
\times 10^{36}$ erg s$^{-1}$ from 5--100 keV, assuming the source
distance to be 7.2 kpc.
 
If the presence of  a cyclotron  resonance scattering feature with $E_{\rm cyc}
\sim22$ keV is confirmed, the magnetic field will be  
$\sim$2.5$\times10^{12}$ G, assuming a 1.4 M$_\odot$ NS with a radius of 10
km and scattering in the accretion column region above the polar cap
on the NS surface.  The  magnetic field of the NS
is given by $E_{\rm cyc} = 11.6~ B_{12}~  (1+z)^{-1}$ keV, where
$B_{12}$ is the  magnetic field in units of 10$^{12}$ G, and $(1+z)$ is the
 gravitational redshift. This value is
consistent for X-ray pulsars with cut-off energies around 10
keV. Unfortunately, the statistics of the high energy
data (with our exposure time) prevent us from confirming the existence
of the CRSF. More extensive  {\em INTEGRAL} observations at high
energy should allow this result to be tested with a longer exposure time.

\acknowledgements
EB is a Chateaubriand Fellow at the Observatory of Paris.
MF acknowledges the French Space Agency and CNRS  for financial
support. The authors are grateful to A. Gros and S. Chazalmartin for
providing the light curve software.
\bibliographystyle{aa}                       
\bibliography{xray}

\end{document}